\documentclass[12pt]{article}
\begin{document}
\title{Resistivity and thermoelectric power measurements on CeFe$_2$ and its pseudobinaries
}
\author{\footnote{corresponding author, e-mail: maulindu@cat.ernet.in, Phone: 91-0731-488348, FAX: 91-0731-488300}  M. K. Chattopadhyay, Meghmalhar Manekar, Kanwal Jeet Singh,\\
 Sujeet Chaudhary, S. B. Roy and P. Chaddah\\
Low Temperature Physics Laboratory,\\
Centre for Advanced Technology,\\ Indore 452013, India}

\date{\today}
\maketitle
\begin{abstract}
Resistivity and thermoelectric power (TEP) measurements on
CeFe$_2$ and two of its pseudo-binaries Ce(Fe, 5\% Ir)$_2$ and
Ce(Fe, 7\% Ru)$_2$ between 78K and 275K are reported. The
resistivity data are analysed in terms of contributions from
scattering due to phonon, magnon, spin fluctuation and lattice
defects, and also from interband scattering. Attempts are made
to analyze the TEP data in terms of these resistivity
components. Thermal hysteresis is observed in the temperature dependence of TEP
in the Ir and Ru doped CeFe$_2$  samples
around the ferromagnetic to antiferromagnetic transition, 
indicating the first order nature of this transition.
\end{abstract}



\newpage

\section{Introduction}
Rare earth-transition metal Laves phase compounds have been under 
intensive theoretical and experimental study in recent years 
because of interesting relationship between magnetism and structure 
in these compounds \cite{1,2,3,4,5}. Among the C15 Laves phase compounds, 
CeFe$_2$ is particularly interesting. It exhibits anomalously low
ferromagnetic (FM) ordering temperature T$_C$ ($<$ 230K), low magnetic
moment per formula unit ($\sim$2.4 $\mu_B$) and smaller lattice
constants compared to other isostructural compounds \cite{6,7,8,9,10,11,12}.
Study of doped CeFe$_2$  has shown that the ferromagnetism 
of CeFe$_2$ is quite fragile
in nature, and a stable low temperature antiferromagnetic (AFM) state can be established
easily with small amount of doping \cite{13,14,15,16,17,18,19,20,21}.

Several reports have been published on the
resistivity \cite{15,16,19,22,36,37}, and some on TEP \cite{36,37} of
CeFe$_2$ and its pseudo-binaries. However, 
various features of resistivity inside different stable magnetic phases  remain not so well understood. 
For example, the sublinear behaviour observed in the resistivity of
CeFe$_2$ and related compounds at high temperatures, 
especially the distinct negative curvature
in the FM regime appears to be quite interesting but any detail 
analysis of these transport properties is lacking so far. In
this paper, we report the results of resistivity and TEP
measurements on CeFe$_2$ and two of its pseudo-binaries Ce(Fe,
5\% Ir)$_2$ and Ce(Fe, 7\% Ru)$_2$ highlighting various
interesting features. We specially focus on the FM
regime, and present an analysis of the data in
terms of different contributions originating due to phonon, magnon, spin fluctuation and
impurity scattering.

\section{Experimental}
The samples used in the present work have also been used earlier in various other
studies \cite{9,16,19,20,22,23}.
Details of sample preparation, heat treatment and characterization 
can be found in Ref.16. 

Resistivity [$\rho(T)$] has been measured by ac technique in the standard four-probe 
configuration, with the help of a SR830DSP lock-in-amplifier coupled to
a SR550 pre-amplifier. Temperature dependence of TEP between 80K and 250K has been measured by a dc differential technique. A temperature difference of $\sim$1K has been maintained across the two ends of the sample. A calibrated 
copper-constantan differential thermocouple has been used to measure
this temperature difference. Thermoelectric voltage has been measured by a Keithley (model 182) sensitive digital voltmeter. The temperature
of the sample has been varied at the rate of 0.3K to 0.4K per minute. Data has been recorded both during heating and cooling to observe the effect of thermal
history on the TEP of the sample. 

\section{Results and Discussion}
Both the $\rho(T)$ and TEP data of CeFe$_2$ and two of its pseudo-binaries Ce(Fe, 5\% Ir)$_2$ and Ce(Fe, 7\% Ru)$_2$ exhibit distinct change of slope at T$_C$, as is evident from 
Figs. 1 and 2 respectively. In the Ir and Ru doped CeFe$_2$ samples, both $\rho(T)$ and TEP
rise with the onset of the lower temperature FM-AFM transition at T$_N$. The paramagnetic
(PM)-FM and FM-AFM transition temperatures tally nicely in the $\rho(T)$ and TEP data, and 
these are also in consonance with other measurements reported earlier \cite{9,19,16,20,22,23}. 
Temperature dependence of TEP shows a distinct thermal hysteresis of width $\sim$6K across 
the FM-AFM transition. No such hysteresis in TEP is observed in any other temperature range including the PM-FM transition regime. 

Our initial attempt to analyse the $\rho(T)$ data using the expression
\begin{eqnarray}
\rho(T) = \rho_0 + \rho_{ph}(T) + \rho_M(T)
\end{eqnarray}
where, $\rho_{ph}$ is given by the Bloch-Gruneisen formula \cite{24}, $\rho_M$ is the resistivity due to magnon scattering as formulated by Fert \cite{25}, and $\rho_0$ is the residual resistivity, did not yield 
good results. Evidently, there are some other contributions to $\rho(T)$ that need to be considered in such analysis.

Paolasini et. al \cite{26} in their inelastic neutron scattering
experiments detected AFM fluctuations contributed by Fe in pure CeFe$_2$, and estimated 
a moment of $\sim$0.05 $\mu_B$ associated with the AFM fluctuations of the Fe
atoms. We argue that such spin fluctuations are likely to contribute to the magnetic scattering 
process of conduction electrons in CeFe$_2$ and related compounds in addition to the standard magnon scattering. To take this into 
account, we add a term $\rho_{sf}$ introduced by Kaiser and Doniach \cite{27}, to 
equation (1) which has been quite successful in analysing the low temperature resistivity 
of a wide variety of materials \cite{28} showing signatures of spin fluctuations. This 
component of resistivity is expressed as:
\begin{eqnarray}
\rho_{sf} = R_s.\Biggl[\frac{\pi}{2}.\Biggl(\frac{T}{T_s}\Biggr) - \frac{1}{2} + \frac{T_s}{4\pi T}.\Psi^{\prime}\Biggl(1 + \frac{T_s}{2\pi T} \Biggr)\Biggr] 
\end{eqnarray}
where, $\Psi^{\prime}(x)$ is the trigamma function, $T_s$ is the spin fluctuation 
temperature, and $R_s$ is a normalization factor depending on the electron-spin fluctuation coupling and on the electronic parameters for the material concerned.

However even the addition of spin fluctuation term to the total resistivity was not adequate enough. We could quantify our results on temperature
dependence of resistivity only after considering an additional $-AT^2$ contribution to 
resistivity, (`$A$' being a constant) originating from impurity scattering into the
$d$-band in these materials, as is explained by Rossiter \cite{30}. Current is largely carried by $s$-electrons, while the $d$-electrons have much higher effective mass and hence much lower mobility. Impurities, phonons and electron-electron interactions can cause scattering of these $s$-electrons into vacant $s$- and $d$-states. But since the scattering probability depends upon the density of states into which the electrons are scattered, $s-d$ scattering can occur much more frequently than $s-s$ scattering. A rapid change in the density of states in the $d$-band, $N_{d}(E_{F})$, with increasing energy can thus lead to a modification in the temperature dependence of resistivity. This is because a thermal broadening of Fermi surface of $\sim kT$ can then produce a significant change in $N_{d}(E_{F})$. It has been shown \cite{30} that such an effect would lead to an additional temperature dependent term of the form $-AT^2$, $A$ being a function of $N(E_{F})$, $dN(E_{F})/dE$, and $d^2N(E_{F})/dE^2$. Such a mechanism has been used \cite{30} to
explain the resistivity of transition metals, which falls below
the linear variation with temperature expected in simple metals
at high temperatures. Thus,
\begin{eqnarray}
\rho(T) = \rho_0 + \rho_{ph}(T) + \rho_m(T) + \rho_{sf}(T) - AT^2
\end{eqnarray}
We used this expression for fitting the data on resistivity in the
FM regime. The various constant terms involved in equation (3), obtained as the
fittings parameters, are shown in Table 1. We assumed $\theta_D$ = 210K
for all the samples according to the specific heat measurement
reports \cite{29}. Once these parameters were obtained, we could
calculate, the exact values of $\rho_{ph}$ \cite{24} and $\rho_{sf}$ \cite{27}
for temperatures beyond the FM regime ($T > T_C$). We then subtracted out the values of$\rho_{ph}$, $\rho_{sf}$, 
$\rho_0$, and $-AT^2$ in the PM and FM regime from the experimental values of resistivity and obtained 
$\rho_m$ as the remainder in the same temperature regime. Fig. 3(a)-(c) show the plots of $\rho_{ph}$, $\rho_{sf}$,
and $\rho_m$  as functions of temperature along with $\rho(T)$ for all the three
samples. It is observed that $\rho_m$, which here denotes the magnetic contribution to 
resistivity for all temperatures, has a distinct change of slope at T$_C$. Below T$_C$, 
$\rho_m$ denotes resistivity due to magnon scattering ($\rho_M$) as usual. So it becomes incoherent at lower temperatures and goes as $T^2$ at higher temperatures (below T$_C$), as is explained by Fert \cite{25}. Because of this $T^2$ dependence, the magnetic component of resistivity becomes quite high in our samples. But we do not compare it with conventional ferromagnets, as the situation in the present samples is quite complicated with the Ce-$4f$ electrons being itinerent, and the nature of their contribution towards electron-electron interaction (which might also go as a $T^2$ term in resistivity) not known completely. At this stage we would
like to point out that the fittings parameters, which are seven in number, can be varied 
up to 10\% to get different combinations that can give good fit between experimental and
calculated values with tolerance less than the error involved in the measurement of 
resistivity. This, however, does not alter the qualitative features of the components, or the gross outcome of the fittings. From Fig. 3 it is clearly observed that the nature
of variation of $\rho_m(T)$ undergoes a marked change due to Ir and Ru
substitution in the pure compound. The contribution of $\rho_{sf}(T)$ is lowest 
in the pure compound, and so is the spin fluctuation temperature (see Table 1). Both are
higher in the Ru doped sample, and for the Ir doped sample they
are the highest. These observations, as we explain below, appear
to be in harmony with the results published by Paolasini et.
al \cite{26}. They found the correlations of AFM fluctuations to vary from
$\sim400 \AA$ at T $<$ 25K to about half of this value at 60K. The AFM fluctuations
reduce in correlation length and increase in frequency with the
rise of temperature and Paolasini et. al expected them to be
observable in careful Mossbauer experiments at temperatures
higher than 60K in the case of pure CeFe$_2$. They imagined a stable FM ground state 
upon which the AFM fluctuations (that have preference for reaching a stable AFM ground state) are formed. We argue from our findings that Ir and Ru doping enhances these AFM fluctuations in terms of 
correlation length, and the peak position in the corresponding spectral density
function (which is the definition of the spin fluctuation temperature $T_s$ in the Kaiser-Doniach \cite{27,28} theory of spin fluctuations; see table 1) also gets shifted to higher temperature. The correlation length becomes much 
larger at lower temperatures. Below a certain characteristic temperature, the FM state
is destroyed completely and a stable AFM state is formed thereafter. 

Though the Kaiser-Doniach expression for $\rho_{sf}$ has been found to be suitable 
for a wide variety of samples \cite{28}, it might give overestimated values at higher
temperatures since it was obtained with random phase approximation which is valid only
in the low temperature limit. In this respect, the theory by Rivier-Zlatic \cite{31} 
is expected to give a better result at high temperatures \cite{32}. According to this
theory, 
\begin{eqnarray}
\rho_{sf} = R_s.\Biggl[1-\Biggl[1+\frac{\pi T}{T_s}+\psi\Biggl(\frac{1}{2}+\frac{T_s}{2 \pi T}\Biggr) - \psi\Biggl(1+\frac{T_s}{2 \pi T}\Biggr)\Biggr]^{-1}\Biggr] 
\end{eqnarray}
Here, $\psi(x)$ denotes digamma function. The fittings parameters for this case is shown 
in Table 2. The components of $\rho(T)$ calculated from these values are not markedly 
different from those obtained employing the Kaiser-Doniach expression for $\rho_{sf}$. 
Thus any of these two theories can be probably used to investigate the present experimental results. We have preferred to continue our analysis using Kaiser-Doniach expression as the Rivier-Zlatic expression has generally been used for the Kondo systems \cite{32,33}.

In our analysis of the temperature dependence of resistivity in Ru and Ir-doped alloys, 
we confined ourselves to the FM regime only. This is because of the lack of proper 
theoretical formulations across the FM-AFM transition. As a result, we had a narrow temperature window of about 25-30K for curve-fitting in the case of the Ce(Fe, 5\% Ir)$_2$ and Ce(Fe, 7\% Ru)$_2$ samples. But $\rho(T)$ measurements in the CeFe$_2$ sample provided us with sufficient data for this purpose (Fig. 4(a)) as we had a wide temperature window extending over $\sim$120K above 78K in which the sample is FM. However, to test our fittings procedure, we have analyzed the data for Ce(Fe, 1\% Ir)$_2$, for which we had data down to 4.2K. These data were obtained earlier in a different set of experiments performed by one of the authors (SBR) on samples of the same batch. This sample did not show any signature for the FM-AFM transition (at least up to temperatures as low as 4.2K). Quite clearly, the data could be fitted reasonably well within the framework described above for the FM state in a wide ($\sim$196K) temperature regime down to 4.2K (Fig. 4(b)).

We now present the results of our analysis of the TEP data. Although the TEP of rare 
earth based intermetallic compounds has often been expressed \cite{34} with the help of the simple Mott formula \cite{35}, it is however unlikely to give quantitatively correct values for TEP as it assumes that the scattering systems are in thermal equilibrium in spite of the presence of the temperature gradient. And in our samples, we have additional sources (other than phonon and magnon) contributing to scattering. Exact theoretical expressions for the contributions of all these sources to TEP are yet to come up. We therefore tried to quantify our TEP data in terms of a modified form of the same Mott formula as,
\begin{eqnarray}
S(T) = A + BT.\Biggl[C + r_1.\frac{\rho_{ph}}{\rho} + r_2.\frac{\rho_m}{\rho} + r_3.\frac{\rho_{sf}}{\rho} + r_4.\frac{\rho_{-AT^2}}{\rho} \Biggr]
\end{eqnarray}
where, $A$, $B$, $C$, $r_1$, $r_2$, $r_3$, and $r_4$ are temperature independent constants. The poor fit to the experimental TEP data obtained using equation (5) (Fig. 2), we believe, is possibly due to the fact that this equation may not exactly represent all the physical processes producing the observed the temperature dependence of TEP.
Nevertheless, the fittings definitely emphasize that the
physical phenomena that give rise to the resistivity components
have an important role to play in the temperature dependence of
TEP as well. The temperature independent term, which comes out
to be 1.97 $\mu$V/K for pure CeFe$_2$ and 1.7 $\mu$V/K and 4.0
$\mu$V/K respectively for the Ru and Ir doped samples, can be 
because of the presence of magnetic impurities in the sample \cite{35}. But the very small 
impurity content in the present samples \cite{16,18,19}, is unlikely to contribute such a 
large value of TEP. The phonon drag TEP is known to be proportional to lattice specific 
heat, and hence to vary as $T^3$ at $T\leq\theta_D/5$. At higher temperatures, 
this contribution is expected to be independent of temperature. But as a 
result of a $T^{-1}$ variation of phonon-phonon scattering relaxation 
time, the phonon drag TEP shows a $T^{-1}$ behaviour at $T > \theta_D$ in 
many materials \cite{35}. However, around $\theta_D$, where the temperature variation of specific heat is negligible, and the phonon relaxation time due to phonon-phonon scattering is nearly independent of temperature, we expect the temperature variation of thermopower to be very slow. We speculate that this contribution can add up with that due to the possible magnetic impurities \cite{16,18,19} to yield a considerably large temperature independent term for thermopower.

The PM-FM transition produces a sharp change of slope at T$_C$ in both $\rho(T)$ and TEP. 
This is in contrast to some of the previous reports \cite{36,37}, wherein the TEP data of 
some members of the CeFe$_2$ family did not show any distinct signature at T$_C$. This 
sharp change of slope in $\rho(T)$ and TEP appearing at the onset of ferromagnetism is 
thought to be due to reduction in spin disorder scattering. Further, the change of
slope in our TEP data resembles that of the transition metals \cite{38}. In contradiction to some earlier reports \cite{37}, no thermal hysteresis of TEP was observed between the 
heating and cooling curves for our samples at and around T$_C$. This absence of hysteresis at T$_C$ is actually in harmony with the second order nature of the PM-FM transition \cite{20}. 

Resistivity and TEP undergo sharp rise in magnitude with lowering of temperature 
at the onset of FM-AFM transition in the Ir and Ru doped compounds. This is attributed 
to the formation of magnetic superzones that deforms a part of the Fermi surface, and
reduces the effective freedom of the conduction electrons \cite{15,16,19}. In the
present samples, the change in TEP across this FM-AFM transition appears to be more 
drastic than that of resistivity. It is known\cite{10} that there is a lattice distortion 
accompanying the FM-AFM transition in the the present compounds. TEP depends directly on the energy derivatives of electron density of states ({\it dN/dE}) and of the collision time ({\it d$\tau$/dE}), which can be quite sensitive to lattice distortions. Hence significant effects might appear in TEP near such transitions \cite{34}. $\rho(T)$, on the other hand, depends primarily on {\it N(E)} and {\it $\tau$(E)} and not on their energy derivatives. Therefore TEP appears to be somewhat more sensitive to the present FM-AFM transition in comparison with the $\rho(T)$ data. The hysteresis between the heating and cooling TEP data across the FM-AFM transition, shown in Fig. 5, is a natural consequence of the first order nature of the transition \cite{20,23}.

\section{Conclusion}
We have investigated the resistivity and thermo electric power of CeFe$_2$ and two of its
pseudo-binaries, Ce(Fe, 5\% Ir)$_2$ and Ce(Fe, 7\% Ru)$_2$. 
FM ordering produces a change of slope in the measured quantities 
across the PM-FM transition. Formation of superzone boundaries at the onset of AFM
ordering causes a remapping of Fermi surface which produces a large change in 
$\rho(T)$. The even more drastic change in TEP across this transition is attributed to
the sensitivity of the energy derivatives of electron density of states and collision time
to the lattice distortion which accompanies the FM-AFM transition. Thermal hysteresis in 
TEP across the FM-AFM transition in Ce(Fe, 5\%Ir)$_2$ and Ce(Fe, 7\% Ru)$_2$ underlines 
the first order character of the transition. Further, $\rho(T)$ of the FM state has been 
analyzed for the first time (to our knowledge) in terms of contributions from scattering 
due to phonon, magnon, spin fluctuations and impurities and the same components have been
used to analyze the TEP data. Last, but not the least, we have highlighted the importance 
of interband scattering effect to explain the interesting resistivity data in the FM regime of CeFe$_2$.

\newpage
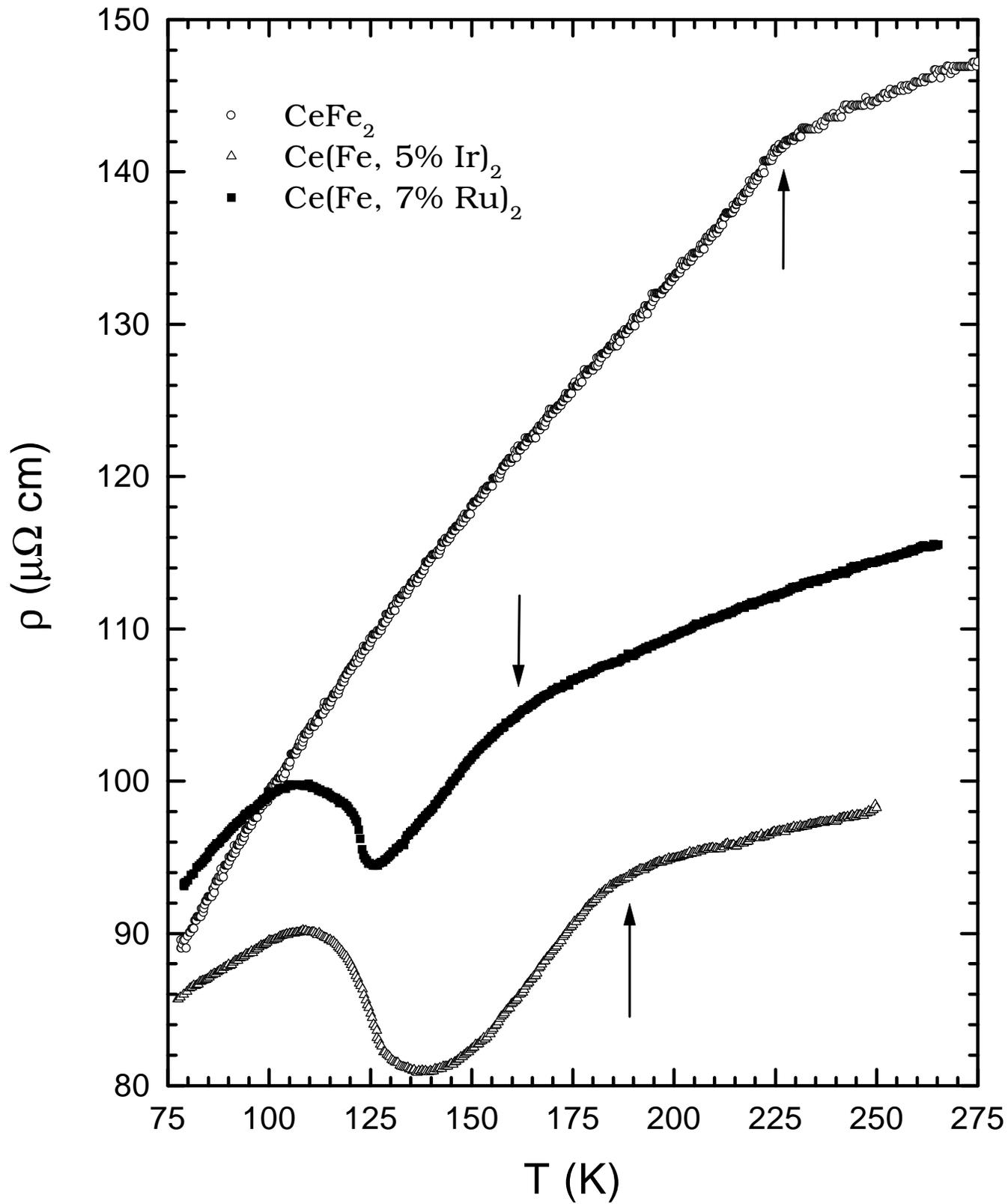
\begin{figure}
\caption{Temperature dependence of resistivity. The arrow-heads indicate T$_C$'s of the respective samples. Increase of resistivity with lowering of temperature represents the FM-AFM transition.} 
\end{figure}
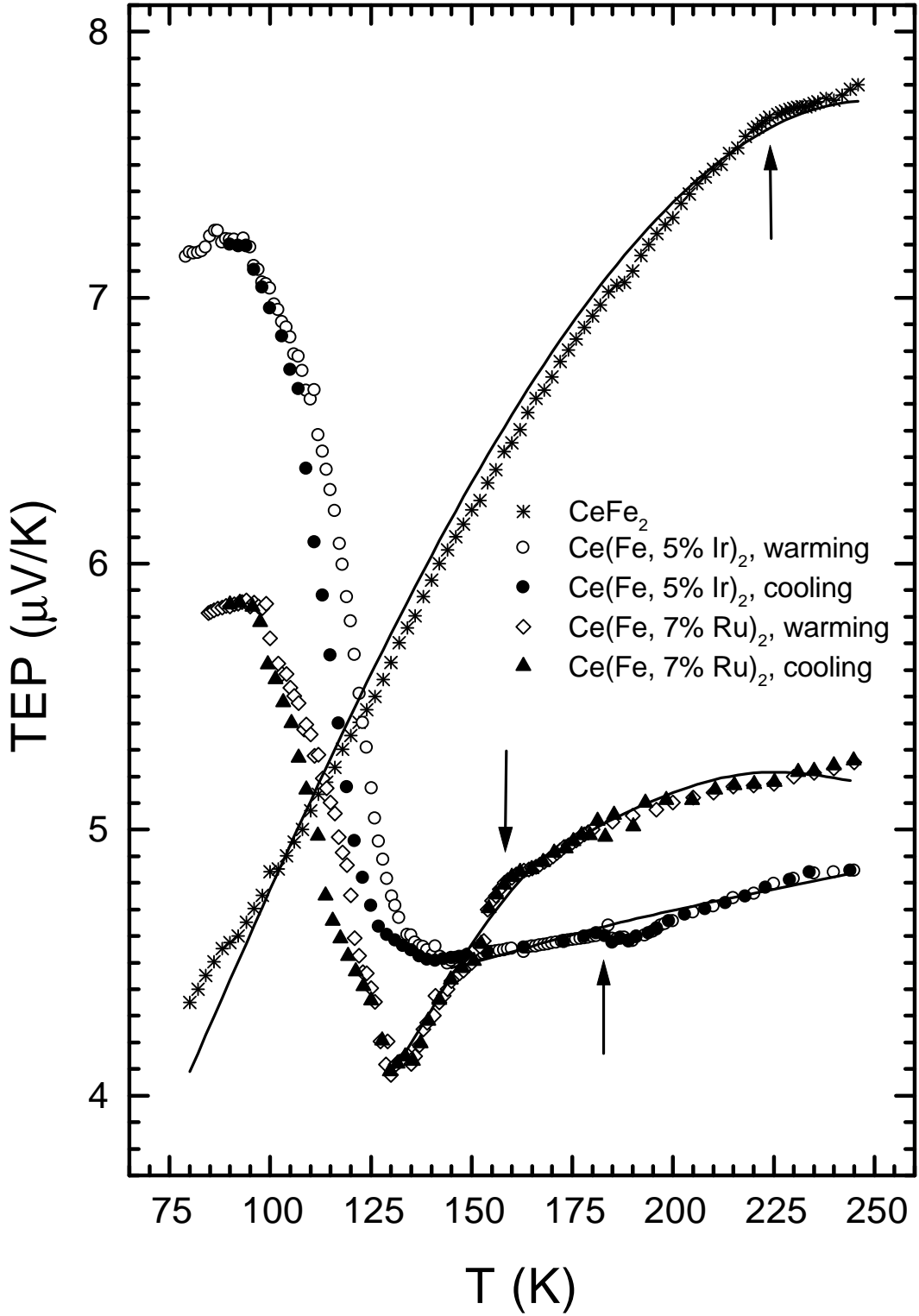
\begin{figure}
\caption{Temperature dependence of thermoelectric power. The arrow-heads indicate T$_C$. The solid line represents equation (5) which is fitted only above the AFM-FM transition.}
\end{figure}
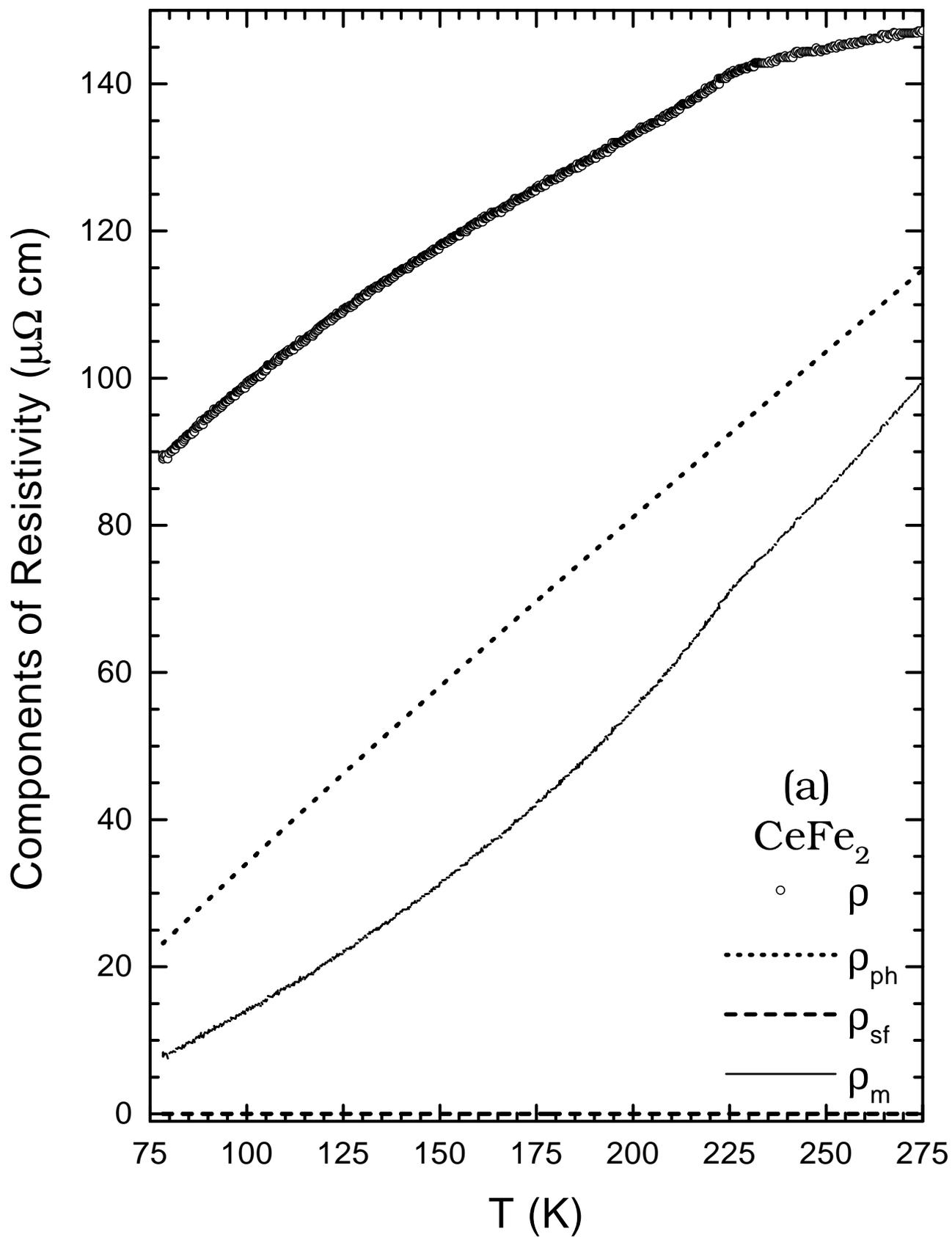
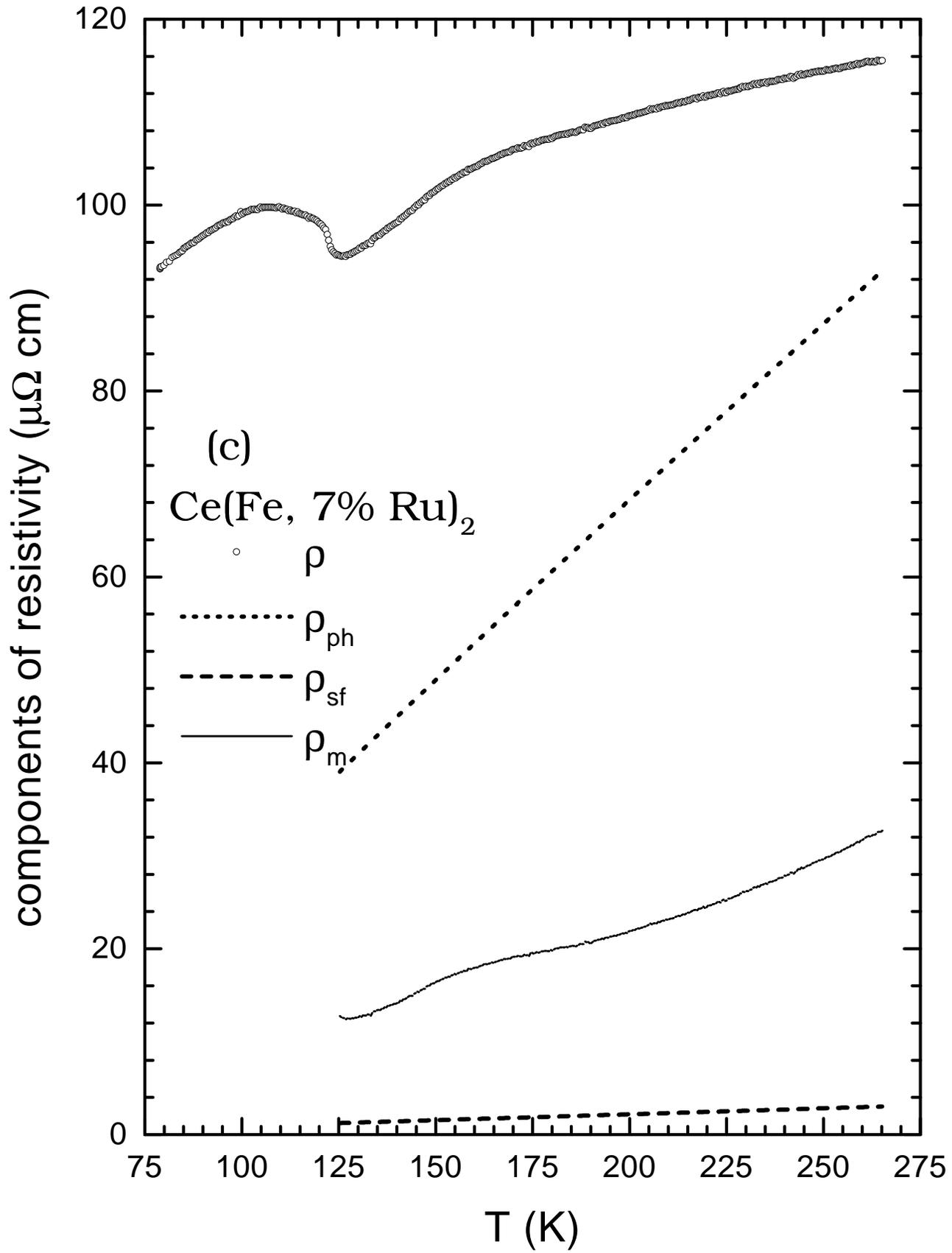
\begin{figure}
\caption{Temperature dependence of the measured resistivity along with its components calculated according to equation (3).}
\end{figure}
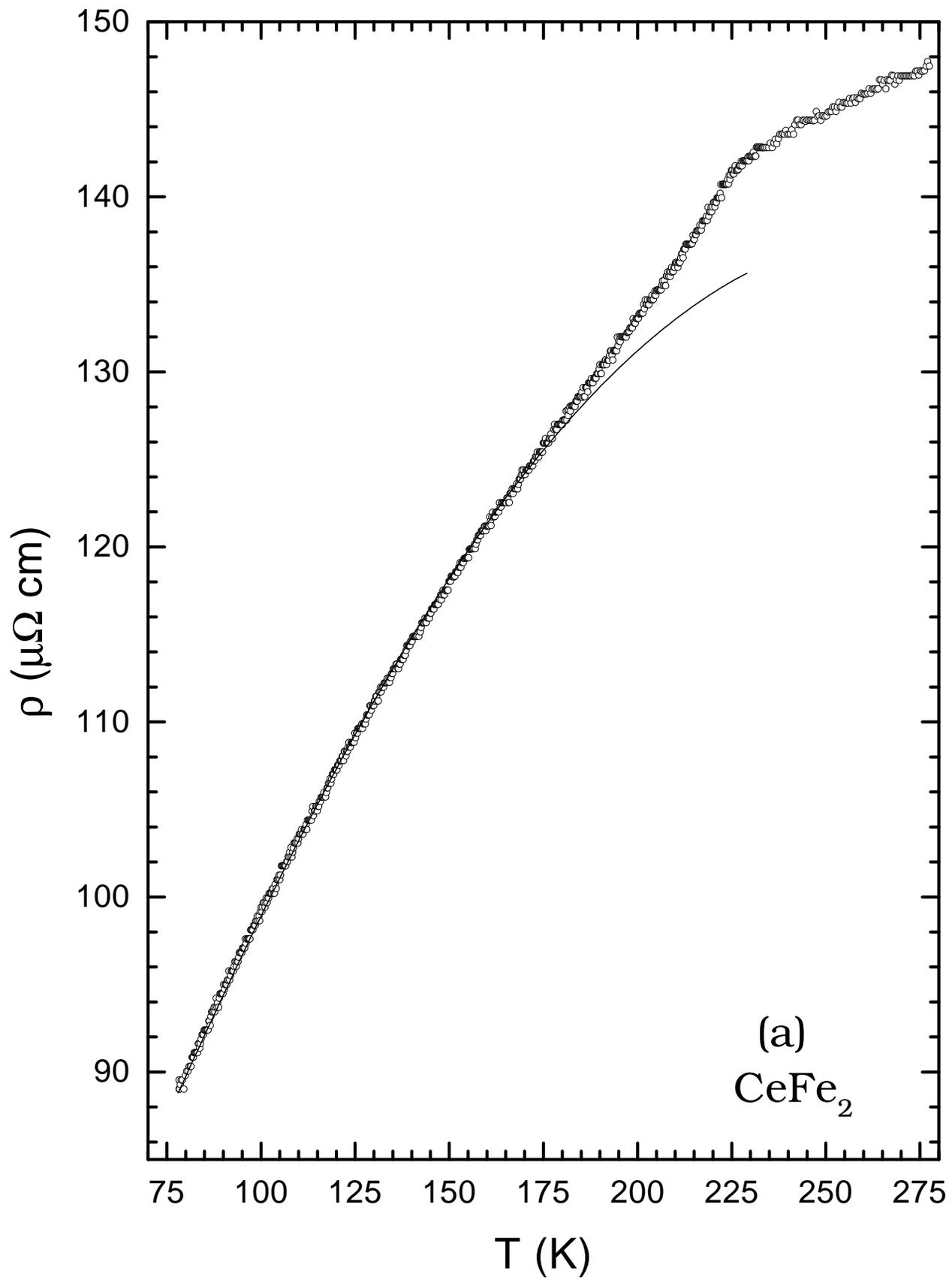
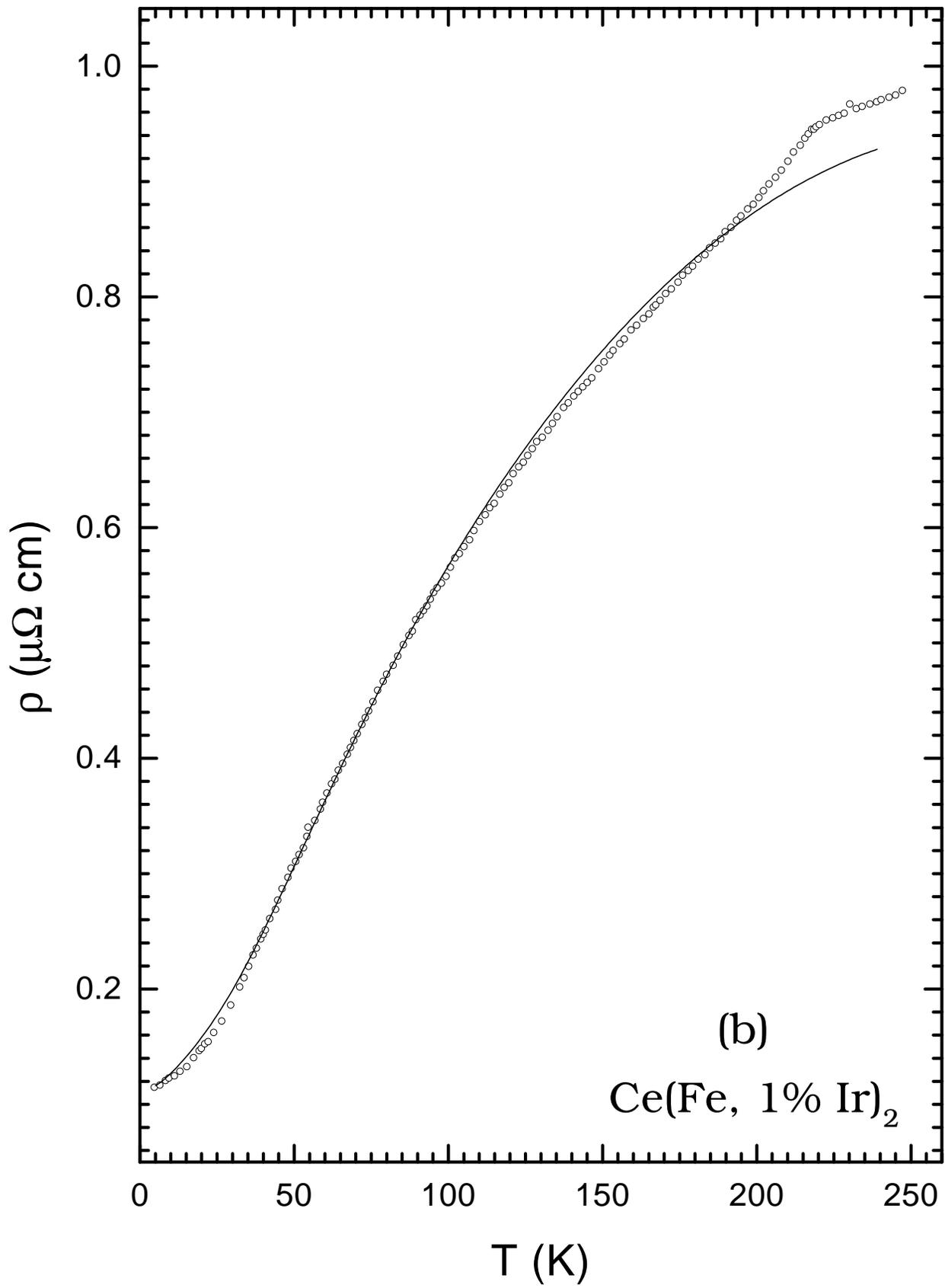
\begin{figure}
\caption{Temperature dependence of resistivity of (a) CeFe$_2$ and (b) Ce(Fe, 1\% Ir)$_2$. The solid lines represent equation (3).} 
\end{figure}
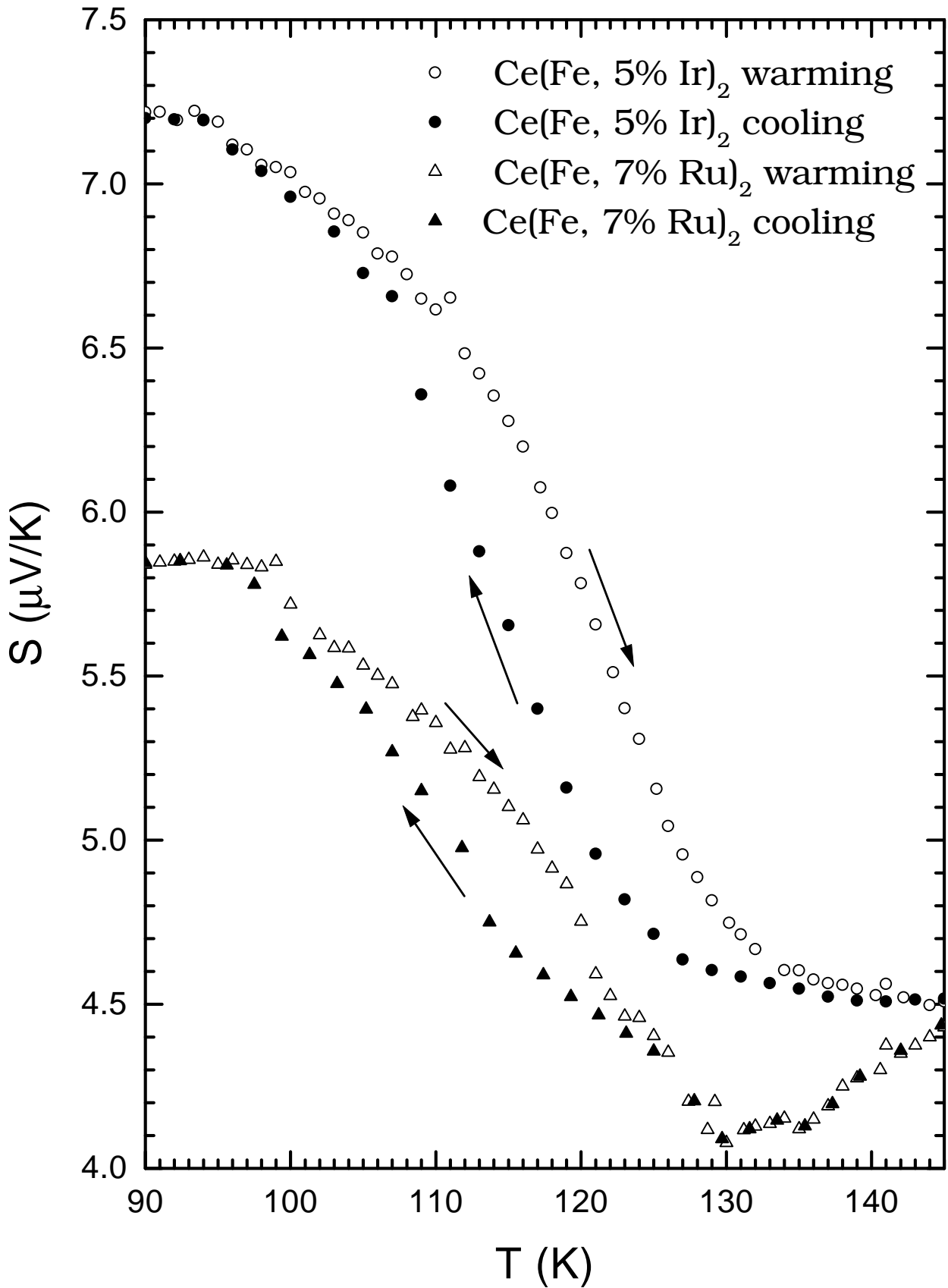
\begin{figure}
\caption{Thermal hysteresis in the thermoelectric power of Ce(Fe, 5\% Ir)$_2$ and Ce(Fe, 7\% Ru)$_2$ in the temperature range around the FM-AFM transition.}
\end{figure}

\begin{table}
\caption{Parameters for equation (3) (using Kaiser-Doniach expression)}
$R_{ph}$ {\it and} $R_M$ {\it are constants associated with phonon and magnon scatterig, and} $\theta_M$ {\it is the characteristic temperature of the magnons.}
\label{table1}
\begin{tabular}{cccc} 
\hline
Parameter&CeFe$_2$&Ce(Fe,5\%Ir)$_2$&Ce(Fe,7\%Ru)$_2$\\  
\hline
$\theta_M$&150.0K&150.0K&150.0K \\ 
$T_s$&70.0K&140.0K&120.0K \\ 
$\rho_0$&69.0 $\mu\Omega$ cm&31.5 $\mu\Omega$ cm&57.2 $\mu\Omega$ cm\\ 
$R_{ph}$&19.0 m$\Omega$ cm K&16.0 m$\Omega$ cm K&16.0 m$\Omega$ cm K\\ 
$R_M$&1.35$\times10^{-3} \mu\Omega$ cm K$^{-2}$&0.75$\times10^{-3} \mu\Omega$ cm K$^{-2}$&0.70$\times10^{-3} \mu\Omega$ cm K$^{-2}$\\ 
$R_s$&5.0$\times10^{-3} \mu\Omega$ cm&5.0 $\mu\Omega$ cm&1.0 $\mu\Omega$ cm\\ 
$A$&1.8$\times10^{-3} \mu\Omega$ cm K$^{-2}$&1.0$\times10^{-3} \mu\Omega$ cm K$^{-2}$&1.0$\times10^{-3} \mu\Omega$ cm K$^{-2}$\\
\hline
\end{tabular}
\end{table}

\begin{table}
\caption{Parameters for equation (3) (using Rivier-Zlatic expression)}
\label{table2}
\begin{tabular}{cccc} 
\hline
Parameter&CeFe$_2$&Ce(Fe,5\%Ir)$_2$&Ce(Fe,7\%Ru)$_2$\\  
\hline
$\theta_M$&150.0K&150.0K&150.0K \\ 
$T_s$&70.0K&140.0K&120.0K \\ 
$\rho_0$&70.0 $\mu\Omega$ cm&38.7 $\mu\Omega$ cm&60.4 $\mu\Omega$ cm\\ 
$R_{ph}$&16.98 m$\Omega$ cm K&14.0 m$\Omega$ cm K&14.0 m$\Omega$ cm K\\ 
$R_M$&1.34$\times10^{-3} \mu\Omega$ cm K$^{-2}$&0.6$\times10^{-3} \mu\Omega$ cm K$^{-2}$&0.70$\times10^{-3} \mu\Omega$ cm K$^{-2}$\\ 
$R_s$&4.0$\times10^{-3} \mu\Omega$ cm&0.1 $\mu\Omega$ cm&0.01 $\mu\Omega$ cm\\ 
$A$&1.545$\times10^{-3} \mu\Omega$ cm K$^{-2}$&0.6$\times10^{-3} \mu\Omega$ cm K$^{-2}$&0.8$\times10^{-3} \mu\Omega$ cm K$^{-2}$\\
\hline
\end{tabular}
\end{table}


\begin{thebibliography}{50}
\bibitem{1}K. H. J. Buschow, in Ferromagnetic Materials, edited by E. P. Wohlfarth (North-Holland, Amsterdam, 1980) Vol. 1, p.297.
\bibitem{2}K. Ikeda, T. Nakamichi, T. Yamada, M. Yamamoto, J. Phys. Soc. Jpn. 36 (1974) 611.
\bibitem{3}S. G. Sankar, W. E. Wallace, in {\it Magnetism and Magnetic Materials (Philadelphia, 1975)} Proceedings of 21 Annual Conference on Magnetism and Magnetic Materials, AIP Conf. Proc. No. 29, edited by J. J. Becker, G. H. Lander, J. J. Rhyne (AIP, New York, 1976) p. 334.
\bibitem{4}T. Nakamichi, J. Phys. Soc. Jpn. 25 (1988) 1189.
\bibitem{5}Y. Yamada, A. Sakata, J. Phys. 57 (1988) 46.
\bibitem{6}J. Deportes, D. Givord, K. R. A. Ziebeck, J. Appl. Phys. 52 (1981) 2074.
\bibitem{7}F. Grandjean, G. D. Waddill, T. R. Cummins, D. P. Moore, G. J. Long, K. H. J. Buschow, Solid St. Commun. 108 (1998) 593.
\bibitem{8}O. Eriksson, L. Nordstrom, M. S. S. Brooks, B. Johansson, Phys. Rev. Lett. 60 (1988) 2523.
\bibitem{9}D. Wang, H. P. Kunkel, G. Williams, Phys. Rev. B 51 (1995) 2872.
\bibitem{10}S. J. Kennedy, B. R. Coles, J. Phys.: Condens. Matter 2 (1990) 1213.
\bibitem{11}S. J. Kennedy, P. J. Brown, B. R. Coles, J. Phys.: Condens. Matter 5 (1993) 5169.
\bibitem{12}M. J. Cooper, P. K. Lawson, M. A. G. Dixon, E. Zukowski, D. N. Timms, F. Itoh, H. Sakurai, H. Kawata, Y. Tanaka, M. Ito, Phys. Rev. B 54 (1996) 4068.
\bibitem{13}D. F. Franceschini, S. F. Da. Cunha, J. Magn. Magn. Mater 52 (1985) 280.
\bibitem{14}A. K. Rastogi, A. P. Murani, in {\it Theoretical and experimental Aspects of Valance Fluctuations and Heavy Fermions}, edited by L. C. Gupta and S. K. Malik (Plenum, New York, 1987) p. 437.
\bibitem{15}S. B. Roy, B. R. Coles, J. Phys.: Condens. Matter 1 (1989) 419.
\bibitem{16}S. B. Roy, B. R. Coles, Phys. Rev. B 39 (1989) 9360.
\bibitem{17}A. K. Rustogi, G. Hilscher, E. Gratz, N. Pillmayr, J. Physique Coll. 49 (1988) C8, 277.
\bibitem{18}S. B. Roy, S. J. Kennedy, B. R. Coles, J. Physique Coll. 49 (1988) C8, 271.
\bibitem{19}A. K. Rajarajan, S. B. Roy, P. Chaddah, Phys. Rev. B 56 (1997) 7808, and references therein.
\bibitem{20}M. Manekar, S. B. Roy, P. Chaddah, J. Phys.: Condens. Matter 12 (2000) L409, and references therein.
\bibitem{21}J. Eynon, N. Ali, J. Appl. Phys. 69 (1991) 5063; Y. S. Yang, B. D. Gaulin, J. A. Fernadz-Baca, N. Ali, G. D. Wingnall, J. Appl. Phys. 73 (1993) 6066.
\bibitem{22}H. P. Kunkel, X. Z. Zhou, P. A. Stampe, J. A. Cowen, G. Williams, Phys Rev. B 53 (1996) 15099.
\bibitem{23}M. Manekar, S. Chaudhary, M. K. Chattopadhyay, K. J. Singh, S. B. Roy, P. Chaddah, J. Phys.: Condens. Matter 12 (2000) 9645.
\bibitem{24}J. M. Ziman, {\it Principles of the Theory of Solids} (Cambridge, Great Britain, 1992) p. 223-225.
\bibitem{25}A. Fert, J. Phys. C (Solid St. Phys.) 2 (1969) 1784.
\bibitem{26}L. Paolasini, P. Dervenagas, P. Vulliet, J. P. Sanchez, G. H. Lander, H. Hiess, A. Panchula, P. Canfield, Phys. Rev. B 58 (1998) 12117.
\bibitem{27}A. B. Kaiser, S. Doniach, Int. J. Magn. 1 (1970) 11.
\bibitem{28}A. B. Kaiser, Philosophical Magazine B 65 (1992) 1197.
\bibitem{29}H. Wada, T. Harada, M. Shiga, J. Phys.: Condens. Matter 9 (1997) 9347.
\bibitem{30}P. L. Rossiter, {\it The Electrical Resistivity of Metals and Alloys} (Cambridge, New York, 1991) p. 273.
\bibitem{31}N. Rivier, V. Zlatic, J. Phys. F 2 (1972) L99.
\bibitem{32}Z. Altounian, S. V. Dantu, M. Dikeakos, Phys. Rev. B 49 (1994) 8621.
\bibitem{33}P. L. Rossiter, {\it The Electrical Resistivity of Metals and Alloys} (Cambridge, New York, 1991) p. 339.
\bibitem{34}M. M. Amado, R. P. Pinto, M. E. Braga, J. B. Sousa, P. Morin, J. Magn. Magn. Mater. 153 (1996) 107.
\bibitem{35}F. J. Blatt, P. A. Schroeder, C. L. Foiles, D. Greig, {\it The Thermoelectric Power of Metals} (Plenum, New York, 1976).
\bibitem{36}E. Gratz, E. Bauer, H. Nowtny, A. T. Burkov, M. V. Vedernikov, Solid St. Commun. 69 (1989) 1007.
\bibitem{37}C. S. Garde, J. Ray, G. Chandra, Phys. Rev. B 42 (1990) 8643.
\bibitem{38}G. A. Thomas, K. Levin, R. D. Parks, Phys. Rev. Lett. 29 (972) 1321.
\end{thebibliography}
 
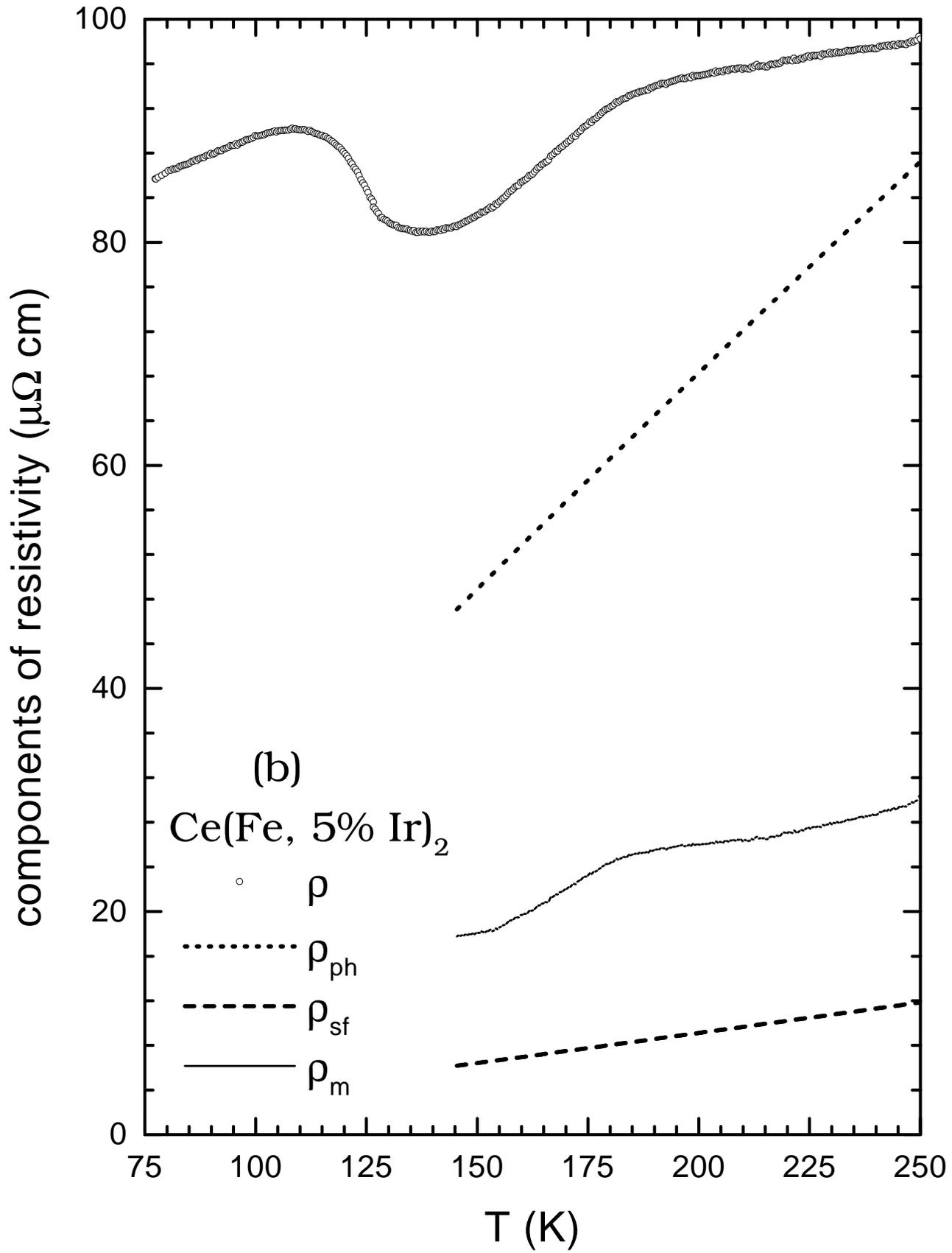
\end{document}